\documentclass[a4shrink,portrait,final]{baposter}

\usepackage{times}
\usepackage{calc}
\usepackage{graphicx}
\usepackage{amsmath}
\usepackage{amssymb}
\usepackage{relsize}
\usepackage{multirow}
\usepackage{bm}

\usepackage{graphicx}
\usepackage{multicol}

\usepackage{pgfbaselayers}
\pgfdeclarelayer{background}
\pgfdeclarelayer{foreground}
\pgfsetlayers{background,main,foreground}

\usepackage{helvet}
\usepackage{bookman}
\usepackage{palatino}

\makeatletter
\def\thebibliography#1{\section*{\@mkboth
  {}{}}\list
  {[\arabic{enumi}]}{\settowidth\labelwidth{[#1]}\leftmargin\labelwidth   
\advance\leftmargin\labelsep
\usecounter{enumi}}
\def\newblock{\hskip .11em plus .33em minus .07em}
\sloppy\clubpenalty4000\widowpenalty4000
\sfcode`\.=1000\relax}
\makeatother

\selectcolormodel{cmyk}

\graphicspath{{images/}}

\newcommand{\ket}[1]{|#1\rangle}

\begin{document}

\title{~~\\~~\\~~\\~~\\~~\\~~\\Quantum Function Computation Using Sublogarithmic Space\thanks{\large This work was partially 
supported by the Scientific and Technological Research Council of Turkey (T\"{U}B\.ITAK) with grant 108142.}}
\author{A. C. Cem Say ~~~~~~ Abuzer Yakary{\i}lmaz 
	\\
	Bo\u{g}azi\c{c}i University, Department of Computer Engineering,\\ Bebek 34342 \.{I}stanbul, Turkey
	\\
	\texttt{ \{say,abuzer\}@boun.edu.tr }
}
\date{December 07, 2009}
\maketitle

\begin{center}

\begin{minipage}{0.70\textwidth}
	\Large
	We examine the power of quantum Turing machines (QTM's) which compute
	string-valued functions of their inputs using small workspace. A (probabilistic or quantum) Turing machine
	has three tapes: A read-only input tape, a read/write work tape, and a
	``1.5-way" write-only output tape. Consider a partial function 
	$ f: \Sigma_{1}^{*}  \rightarrow \Sigma_{2}^{*} $, where $ \Sigma_{1} $ and $ \Sigma_{2} $ 
	are the input and output
	tape alphabets, respectively. We say that such a machine $ \mathcal{M} $ 
	\textit{computes the function f with bounded error}
	if, when started with string $ w \in \Sigma_{1}^{*} $ on its input tape,
	\begin{enumerate}
		\item the probability that $ \mathcal{M} $ halts by arriving at an accept state, having
			written the string $ f(w) $ on its output tape, is at least $ \frac{2}{3} $, if $ f(w) $
			is defined, and,
		\item the probability that $ \mathcal{M} $ halts by arriving at a reject state is at
			least $ \frac{2}{3} $, if $ f(w) $ is undefined.
	\end{enumerate}
	We compare QTM's with probabilistic Turing machines (PTM's) under the
	same space bounds. It is straightforward to show that any function
	computable by a PTM in space $s$ can be computed in space $s$ by a
	QTM, for any space bound $s$. We prove that QTM's are strictly
	superior to PTM's in this regard for any $ s=o(\log n) $, by demonstrating
	that there exists a quantum finite state transducer (QFST), that is, a
	QTM which does not use its work tape, and moves the input head to the
	right in every step of its execution, which computes the function
	\[ 
		f_{1}(x) = \left\lbrace 
			\begin{array}{ll}
				w, & \mbox{ if } x=w2w, \mbox{ where } w \in \{0,1\}^{*}, \\
				\mbox{undefined}, & \mbox{ otherwise}
			\end{array}
		\right.
	\]
	with bounded error. This function cannot be computed with bounded
	error by any PTM using $ o(\log n) $ space.
	
	Our QFST for $ f_1 $ is also interesting as a language recognition device,
	in the sense that it is the first example for a quantum  finite-state
	machine which traverses its input with a one-way head, and is able to
	recognize a nonregular language with bounded error.
\end{minipage}

\end{center}

%%%%%%%%%%%%%%%%%%%%%%%%%%%%%%%%%%%%%%%%%%%%%%%%%%%%%%%%%%%%%%%%%%%%%%%%%%%%%%
%%% Here starts the poster
%%%---------------------------------------------------------------------------
%%% Format it to your taste with the options
%%%%%%%%%%%%%%%%%%%%%%%%%%%%%%%%%%%%%%%%%%%%%%%%%%%%%%%%%%%%%%%%%%%%%%%%%%%%%%
% Define some colors
\definecolor{silver}{cmyk}{0,0,0,0.3}
\definecolor{yellow}{cmyk}{0,0,0.9,0.0}
\definecolor{reddishyellow}{cmyk}{0,0.22,1.0,0.0}
\definecolor{black}{cmyk}{0,0,0.0,1.0}
\definecolor{darkYellow}{cmyk}{0,0,1.0,0.5}
\definecolor{darkSilver}{cmyk}{0,0,0,0.1}

\definecolor{lightyellow}{cmyk}{0,0,0.3,0.0}
\definecolor{lighteryellow}{cmyk}{0,0,0.1,0.0}
\definecolor{lighteryellow}{cmyk}{0,0,0.1,0.0}
\definecolor{lightestyellow}{cmyk}{0,0,0.05,0.0}

\typeout{Poster Starts}
\background{
  \begin{tikzpicture}[remember picture,overlay]%
    \draw (current page.north west)+(-2em,2em) node[anchor=north west] 
    {\includegraphics[height=1.1\textheight]{silhouettes_background}};
  \end{tikzpicture}%
}

\newlength{\leftimgwidth}
\begin{poster}%
  % Poster Options
  {
  % Show grid to help with alignment
  grid=no,
  % Column spacing
  colspacing=1em,
  % Color style
  bgColorOne=lighteryellow,
  bgColorTwo=lightestyellow,
  borderColor=reddishyellow,
  headerColorOne=yellow,
  headerColorTwo=reddishyellow,
  headerFontColor=black,
  boxColorOne=lightyellow,
  boxColorTwo=lighteryellow,
  % Format of textbox
  textborder=roundedleft,
  % Format of text header
  eyecatcher=no,
  headerborder=open,
  headerheight=0.09\textheight,
  headershape=roundedright,
  headershade=plain,
  headerfont=\Large\textsf, %Sans Serif
  boxshade=plain,
  background=shade-tb,
  background=plain,
  linewidth=1pt
  }
  % Eye Catcher
  {\includegraphics[width=10em]{D1077}} 
  % No eye catcher for this poster. (eyecatcher=no above). If an eye catcher is present, 
  %the title is centered between eye-catcher and logo.
  % Title
	{\sf %Sans Serif %\bf% Serif
		Quantum Function Computation Using Sublogarithmic Space
	}
  % Authors
  {\sf \\ A. C. Cem Say and Abuzer Yakary{\i}lmaz \\
  Bo\u{g}azi\c{c}i University, Department of Computer Engineering, Bebek 34342 \.{I}stanbul, Turkey
  }
  % University logo
%  {% The makebox allows the title to flow into the logo, this is a hack because of the L shaped logo.
%    \makebox[8em][r]{%
%      \begin{minipage}{16em}
%        \hfill
%        %\includegraphics[height=2em]{msrlogo}
%        %\includegraphics[height=7.0em]{logo}
%      \end{minipage}
%    }
%  }

 \tikzstyle{light shaded}=[top color=baposterBGtwo!30!white,bottom color=baposterBGone!30!white,shading=axis,shading angle=30]

  % Width of left inset image
   %  \setlength{\leftimgwidth}{0.78em+8.0em}

%%%%%%%%%%%%%%%%%%%%%%%%%%%%%%%%%%%%%%%%%%%%%%%%%%%%%%%%%%%%%%%%%%%%%%%%%%%%%%
%%% Now define the boxes that make up the poster
%%%---------------------------------------------------------------------------
%%% Each box has a name and can be placed absolutely or relatively.
%%% The only inconvenience is that you can only specify a relative position 
%%% towards an already declared box. So if you have a box attached to the 
%%% bottom, one to the top and a third one which should be in between, you 
%%% have to specify the top and bottom boxes before you specify the middle 
%%% box.
%%%%%%%%%%%%%%%%%%%%%%%%%%%%%%%%%%%%%%%%%%%%%%%%%%%%%%%%%%%%%%%%%%%%%%%%%%%%%%
    %
    % A coloured circle useful as a bullet with an adjustably strong filling
    \newcommand{\colouredcircle}[1]{%
      \tikz{\useasboundingbox (-0.2em,-0.32em) rectangle(0.2em,0.32em); \draw[draw=black,fill=baposterBGone!80!black!#1!white,line width=0.03em] (0,0) circle(0.18em);}}
\large{
%%%%%%%%%%%%%%%%%%%%%%%%%%%%%%%%%%%%%%%%%%%%%%%%%%%%%%%%%%%%%%%%%%%%%%%%%%%%%%
  \headerbox{Models}{name=Definition,column=0,row=0,span=3}{
%%%%%%%%%%%%%%%%%%%%%%%%%%%%%%%%%%%%%%%%%%%%%%%%%%%%%%%%%%%%%%%%%%%%%%%%%%%%%%
A (probabilistic or quantum) Turing machine (PTM or QTM) has a finite
register and three tapes:
\begin{enumerate}
       \item A read-only input tape,
       \item a read/write work tape, and
       \item a $ ``\mbox{1.5-way}" $ write-only output tape.
\end{enumerate}
The finite register is observed after each transition in order to
check whether the computation has halted by
accepting or rejecting, or not. The register is refreshed before the
next transition.
This mechanism allows QTM's to implement general quantum operators.
\\[5pt]
A quantum finite state transducer (QFST) is a QTM which does not use
its work tape, and moves the input head
to the right in every step of its execution.
}
%%%%%%%%%%%%%%%%%%%%%%%%%%%%%%%%%%%%%%%%%%%%%%%%%%%%%%%%%%%%%%%%%%%%%%%%%%%%%%
\headerbox{Function Computation}{name=Function-Computation,column=0,below=Definition,span=3}{
%%%%%%%%%%%%%%%%%%%%%%%%%%%%%%%%%%%%%%%%%%%%%%%%%%%%%%%%%%%%%%%%%%%%%%%%%%%%%%
Consider a partial function
$ f: \Sigma_{1}^{*}  \rightarrow \Sigma_{2}^{*} $, where $ \Sigma_{1}
$ and $ \Sigma_{2} $
are the input and output tape alphabets, respectively. We say that a
machine $ \mathcal{M} $
\textit{computes the function $ f $ with bounded error}
if, when started with string $ w \in \Sigma_{1}^{*} $ on its input tape,
\begin{enumerate}
       \item the probability that $ \mathcal{M} $ halts by observing an
accept symbol in the register, having
               written the string $ f(w) $ on its output tape, is at least $
\frac{2}{3} $, if $ f(w) $
               is defined, and,
       \item the probability that $ \mathcal{M} $ halts by observing a
reject symbol in the register is at
               least $ \frac{2}{3} $, if $ f(w) $ is undefined.
\end{enumerate}
}

%%%%%%%%%%%%%%%%%%%%%%%%%%%%%%%%%%%%%%%%%%%%%%%%%%%%%%%%%%%%%%%%%%%%%%%%%%%%%%
  \headerbox{A QFST algorithm}{name=QFST-Algorithm,column=0,below=Function-Computation,span=3}{
%%%%%%%%%%%%%%%%%%%%%%%%%%%%%%%%%%%%%%%%%%%%%%%%%%%%%%%%%%%%%%%%%%%%%%%%%%%%%%
We compute the function
\[
       f_{1}(x) = \left\lbrace
               \begin{array}{ll}
                       w, & \mbox{ if } x=wcw, \mbox{ where } w \in \{a,b\}^{*}, \\
                       \mbox{undefined}, & \mbox{ otherwise}
               \end{array}
       \right.
\]
with bounded error by a QFST as follows:
\begin{enumerate}
       \item The computation splits into three paths,
               $ \mathsf{path}_{1} $, $ \mathsf{path}_{2} $, and $ \mathsf{path}_{3} $,
               with equal probability at the beginning.
       \item $ \mathsf{path}_{3} $ rejects immediately.
       \item $ \mathsf{path}_{1} $ ($ \mathsf{path}_{2} $) outputs $ w_{1} $
($ w_{2} $) if the input is of the form $ w_{1} c w_{2} $,
               where $ w_{1},w_{2} \in \{a,b\}^{*} $.
               \begin{enumerate}
               \item If the input is not of the form $ w_{1} c w_{2} $, both paths reject.
               \item Otherwise, at the end of the computation,
               $ \mathsf{path}_{1} $ and $ \mathsf{path}_{2} $ perform the following QFT:
               \[
                       \begin{array}{lclcl}
                               \mathsf{path}_{1} & \rightarrow & \frac{1}{\sqrt{2}}
                               \ket{q,``\mathtt{Accept}",w_{1}} & +
                               & \frac{1}{\sqrt{2}} \ket{q,``\mathtt{Reject}",w_{1}}  \\
                               \mathsf{path}_{2} & \rightarrow & \frac{1}{\sqrt{2}}
                               \ket{q,``\mathtt{Accept}",w_{2}} & -
                               & \frac{1}{\sqrt{2}} \ket{q,``\mathtt{Reject}",w_{2}}  \\
                       \end{array}
                \]
       \end{enumerate}
\end{enumerate}
\begin{center}
The configurations belonging to $ \mathsf{path}_{1} $ and $ \mathsf{path}_{2} $
interfere with each other, i.e., the machine accepts with probability
$ \frac{2}{3} $,
\end{center}
\begin{center}
if and only if
\end{center}
\begin{center}
the input is of the form $ w c w $, $ w \in \{a,b\}^{*} $, the case
where $ f_{1} $ is defined.
\end{center}

}

%%%%%%%%%%%%%%%%%%%%%%%%%%%%%%%%%%%%%%%%%%%%%%%%%%%%%%%%%%%%%%%%%%%%%%%%%%%%%%
 \headerbox{Main Result}{name=Main-Result,column=0,below=QFST-Algorithm,span=3}{
%%%%%%%%%%%%%%%%%%%%%%%%%%%%%%%%%%%%%%%%%%%%%%%%%%%%%%%%%%%%%%%%%%%%%%%%%%%%%%
\textbf{Fact 1.} Any function computable by a PTM in space $s$ can be
computed in space $s$ by a
QTM, for any space bound $s$.
\\[10pt]
\textbf{Fact 2.} No PTM can recognize $ L_{pal} = \{ w \mid
w=w^{\mathtt{R}}, w \in \{a,b\}^{*} \} $
with bounded error using $ o(\log n) $ space.
\\[10pt]
\textbf{Fact 3.} Function $ f_{1} $ cannot be computed with bounded
error by any PTM using $ o(\log n) $ space.
Otherwise, we would have a PTM recognizing the language
$ \{ wcw \mid w \in \{a,b\}^{*} \} $ and also $ L_{pal} $
with bounded error using $ o(\log n) $ space. This would contradict Fact 2.
\\[10pt]
\Large{
\textbf{Main Result.} We prove that QTM's are strictly superior to
PTM's in function computation for any space
$ o(\log n) $ due to the QFST algorithm and Facts 1 and 3.}
\\[10pt]
\large{
\textbf{Remark.} Our QFST for $ f_1 $ is also interesting as a
language recognition device, in the sense that it is the
first example for a quantum  finite-state machine which traverses its
input with a one-way head, and is able to
recognize a nonregular language with bounded error.
}
}

%%%%%%%%%%%%%%%%%%%%%%%%%%%%%%%%%%%%%%%%%%%%%%%%%%%%%%%%%%%%%%%%%%%%%%%%%%%%%%%
%  \headerbox{References}{name=References,column=0,span=3,below=Main-Result}{
%%%%%%%%%%%%%%%%%%%%%%%%%%%%%%%%%%%%%%%%%%%%%%%%%%%%%%%%%%%%%%%%%%%%%%%%%%%%%%%
%    \smaller
%    \vspace{-14pt}
%    \bibliographystyle{plain}
%	\bibliography{YakaryilmazSay}
%}
} % END of \large

\end{poster}%

\end{document}